\documentclass[11pt,preprint,showpacs]{revtex4}
\usepackage{graphicx}% Include figure files
\usepackage{dcolumn}% Align table columns on decimal point
\usepackage{bm}% bold math
\usepackage{amsmath}
\usepackage{amssymb}
\usepackage{amsfonts}
\def\bsig{\mbox{\boldmath$\sigma$}}
\def\btheta{\mbox{\boldmath$\theta$}}

\def\brho{\mbox{\boldmath$\rho$}}
\begin{document}
%\pagestyle{plain}
%\draft

\title{Relativity and the low energy nd $A_y$ puzzle}

\author{H.~Wita{\l}a}
\affiliation{M. Smoluchowski Institute of Physics, Jagiellonian
University,
                    PL-30059 Krak\'ow, Poland}

\author{J.~Golak}
\affiliation{M. Smoluchowski Institute of Physics, Jagiellonian
University,
                    PL-30059 Krak\'ow, Poland}

\author{R.~Skibi\'nski}
\affiliation{M. Smoluchowski Institute of Physics, Jagiellonian
University,
                    PL-30059 Krak\'ow, Poland}

\author{W.\ Gl\"ockle}
\affiliation{Institut f\"ur theoretische Physik II,
Ruhr-Universit\"at Bochum, D-44780 Bochum, Germany}

\author{W. N.\ Polyzou}
\affiliation{Department of Physics and Astronomy,
The University of Iowa, Iowa City, IA 52242}

\author{H.\ Kamada}
\affiliation{Department of Physics, Faculty of Engineering,
Kyushu Institute of Technology, Kitakyushu 804-8550, Japan}

\date{\today}

\begin{abstract}

We solve the Faddeev equation in an exactly Poincar\'e invariant
formulation of the three-nucleon problem.  The dynamical input is a
relativistic nucleon-nucleon (NN) interaction that is exactly on-shell
equivalent to the high precision 
 CD~Bonn NN interaction.  $S$-matrix cluster
properties dictate how the two-body dynamics is embedded in the
three-nucleon mass operator (rest Hamiltonian).  We find that for
neutron laboratory energies above $\approx 20$~MeV relativistic
effects on $A_y$ are negligible.  For energies below $\approx 20$~MeV
dynamical effects lower the nucleon analyzing power maximum slightly
by $\approx 2\%$ and Wigner rotations lower it further up to $\approx
10 \%$ increasing thus disagreement between data and theory. This
indicates that three-nucleon forces (3NF) must provide an even larger
increase of the $A_y$ maximum than expected up to now.

\end{abstract}


\maketitle
\setcounter{page}{1}

\section{Introduction}

High precision nucleon-nucleon (NN) interactions such as
AV18~\cite{AV18}, CDBonn~\cite{CDBONN}, Nijm I, II and 93~\cite{NIJMI}
accurately describe the NN data set up to about 350 MeV.  When these
interactions are used to predict binding energies of three-nucleon
(3N) systems they underestimate the experimental bindings of $^3H$ and
$^3He$ by about 0.5-1 MeV~\cite{Friar1993,Nogga1997}. This missing
binding energy can be cured by introducing a three-nucleon force (3NF)
into the nuclear Hamiltonian~\cite{Nogga1997}.

The study of elastic nucleon-deuteron (Nd) scattering and
nucleon-induced deuteron breakup also revealed a number of cases where
the nonrelativistic description based on pairwise interactions is
insufficient to explain the data.  Generally, the studied
discrepancies between a theory based on NN interactions only and
experiment become larger with increasing energy of the 3N
system. Adding a 3NF that includes long-range $ 2 \pi$ exchange to the
pairwise interactions leads in some cases to a better description of
the data. The parameters of such a 3NF must be separately adjusted to
the experimental binding of $^3H$ and $^3He$~\cite{wit98,wit01,sek02}
for each NN interaction. The elastic Nd angular distribution in the
region of its minimum and at backward angles is the best studied
example~\cite{wit98,sek02}. The clear discrepancy in these angular
regions at energies below $\approx 100$~MeV nucleon laboratory energy
between a theory based on NN interactions only and the cross-section
data can be removed by adding modern 3NFs based on chiral effective
field theory \cite{epel2002} to the nuclear Hamiltonian.  At energies
higher than $\approx 100$~MeV current 3NFs \cite{TM,uIX} only
partially improve the description of cross section data and the
remaining discrepancies, which increase with energy, indicate the
possibility of relativistic effects~\cite{erm1,erm2,maeda}.  The need
for a relativistic description of 3N scattering was also raised when
precise measurements of the total cross section for neutron-deuteron
(nd) scattering~\cite{abf98} were analyzed within the framework of
nonrelativistic Faddeev calculations~\cite{wit99}.  NN interactions
alone were insufficient to describe the data above $\approx 100$~MeV.

In few-body models off-shell effects, relativistic effects, and
three-body force contributions cannot be cleanly separated.  This is
because different two-body interactions that give the same two-body
$S$ matrix are related by a unitary scattering
equivalence~\cite{ekstein60}.  To maintain this equivalence at the
three-body level requires additional three-body interactions
\cite{wgwp90} in one of the Hamiltonians.  Since relativistic two-body
models are fit to the same data as the corresponding nonrelativistic
models, there is a similar on-shell two-body scattering equivalence.
While the relativistic and nonrelativistic three-body predictions
will be different one can in principle make up the difference (in a
chosen frame) with a suitable three-body interaction.  So while it is
possible to simulate ``relativistic effects'' with a three-body
interaction, a Poincar\'e invariant treatment of the dynamics provides
the most direct way to model the consequences of imposing Poincar\'e
invariance and S-matrix cluster properties.

In this paper we investigate one particular representation of the
Poincar\'e invariant three-body problem.  We compare the predictions
of relativistic and nonrelativistic three-body calculations where the
input two-body interactions give the same two-body S matrix, have the
same internal two-body wave functions, and have a kinematic
three-dimensional Euclidean symmetry. The relativistic and
nonrelativistic models differ in how these interactions appear in the
three body-problem.  In addition, the internal and single particle
variables are related by Galilean boosts in the nonrelativistic case
and Lorentz boosts in the relativistic case.  In the nonrelativistic
case the two-body interactions for each pair are added to the center
of mass kinetic energy operator.  In the relativistic case, the
non-linear relation between the two and three-body mass operators must
be respected in order to obtain a scattering matrix that clusters into
a product of the identity and the input two-body $S$-matrix
\cite{relform1} .
This non-linear dependence of the three-body invariant mass operator
on the two-body interaction has dynamical consequences for the
three-body system which complicates the structure of the Faddeev kernel.

In ~\cite{witrel} we used a Poincar\'e invariant formulation of the 3N
scattering problem.  A technique for constructing the relativistic
nucleon-nucleon interaction from a standard high-precision interaction
was given in ~\cite{kam2002}.  We used the same technique to
construct the transition operators that appear in the kernel of the
relativistic Faddeev equation.  Application to a 3N bound state
supported the relativistic effects previously found in ~\cite{relform}.

Realistic NN interactions are fit by properly transforming
experimental data to the center of momentum frame and fitting
$S$-matrix elements computed using the nonrelativistic Schr\"odinger
equation to this data.  While the same data could be precisely fit
using $S$-matrix elements computed from a relativistic Schr\"odinger
equation, this has not been done \cite{schiavilla} with the same
precision used to construct realistic interactions.

In~\cite{kam2002}, instead, an analytical scale transformation of
momenta was used to relate NN interactions in the nonrelativistic and
relativistic Schr\"odinger equations in such a way, that the 
two-body scattering matrix elements are identified as
$S_{nr}(E_{cm}) = S_{r}(E_{cm})= S_{exp}(E_{cm})$ as functions of 
the center of momentum energy~\cite{kam98}.
In this work we use an alternative procedure \cite{cps75} that generates
a relativistic nucleon-nucleon interaction with the property that the
relativistic and nonrelativistic two-body $S$ matrices satisfy
$S_r(\mathbf{k}^2) = S_{nr}(\mathbf{k}^2)= S_{exp}(\mathbf{k}^2)$,
where $S_{exp}(\mathbf{k}^2)$ is the experimental two-body $S$ matrix,
and $\mathbf{k}^2$ is the cm momentum of one of the particles.

When high-precision potentials are determined \cite{app00} by
properly Lorentz transforming scattering data from the laboratory
frame to the center of momentum frame the Lorentz invariant scalar
product $p_{target} \cdot p_{beam} = m E_b = 2\mathbf{k}^2 + m^2$ is
used to relate the laboratory beam energy $E_b$ to the c.m. momentum
$\mathbf{k}^2$.  The potential is determined by comparing the
transformed experimental scattering observables to the scattering
observables computed using the nonrelativistic Lippmann-Schwinger
equation, identifying the $\mathbf{k}^2$ appearing in the
Lippmann-Schwinger equation with the $\mathbf{k}^2$ computed from the
invariant $p_{target} \cdot p_{beam}$. With this procedure the
resulting interactions are constructed so the S-matrix elements 
in the relativistic and nonrelativistic cases are
identified as functions of $\mathbf{k}^2$ rather than cm energy.
Though the difference in the two approaches leads to a small mismatch
in the relativistic and nonrelativistic momentum, the 
interactions generated by the analytic scale transformation 
provide a useful first step to investigate the effects of
the non-linear relation between the three-body mass operator and
two-body interactions.

In our initial studies~\cite{witrel,witbr,skibbr} 
the interaction generated by the analytic scale transformation
was used
to study the changes in elastic nd scattering and breakup
observables when the nonrelativistic form of the kinetic energy is
replaced by the relativistic one and a proper treatment of the
dynamics is included.  We found that the elastic scattering cross
section is only slightly influenced by relativity.  Only at backward
angles and higher energies the elastic cross sections are increased by
relativity.

Due to the selectivity of the breakup reaction, however,
regions of phase-space were found at higher energies of the incoming
nucleon where relativity leads to a characteristic pattern by which
relativity changes the nonrelativistic breakup cross section.
Namely, in this region of phase-space fixing the angle of the first
detected nucleon and changing the angle of the second nucleon provides
variations of the nonrelativistic cross section by relativity,
increasing or decreasing it by a factor up to $\approx 2$. For spin
observables the implemented relativistic features lead only to small
effects.

Recently an interesting estimate of ``relativistic corrections'' has
been performed and its effect on low-energy nd analyzing power $A_y$
has been estimated using the plane-wave impulse
approximation~\cite{schwenk}.  The estimate is based on a perturbative
realization of the Poincar\'e Lie algebra to leading order in $1/c^2$
\cite{foldy74}.  The calculations are done in the plane-wave impulse
approximation.  A large increase by $\approx 10 \%$ of the $A_y$
maximum at laboratory energy $E_n=3$~MeV has been found.  The authors
comment that their estimates are both exploratory and incomplete.  In
addition to the absence of final state interaction, there are a number
of other important differences with an exact formulation of this
problem.  Such a large effect, which would significantly reduce the
discrepancy between theory and data in the region of the $A_y$ maximum,
calls for a relativistic study in an exactly Poincar\'e invariant
treatment of the three nucleon dynamics.

Our previous study~\cite{witrel}, performed without inclusion of
Wigner rotations, is too limited for spin observables.  Therefore, in
order to make definite conclusions for $A_y$ we perform a complete
dynamical calculation including the effects of Wigner rotations. We
focus on that issue and do not include 3NF's.

The paper is organized as follows.  In section II we discuss the
construction of our Poincar\'e invariant dynamical model.  This includes
a discussion of how high-precision interactions are used to construct
the three-body mass operator (rest Hamiltonian).  In section III we
discuss spin observables in Poincar\'e invariant quantum theory.  In
Sec. IV we discuss the formulation of the Faddeev equation to
construct scattering observables for this three-nucleon mass operator.
This includes an exact treatment of the Faddeev kernel which avoids
the approximations used in~\cite{witrel,witbr,skibbr}.  We solve the
relativistic 3N Faddeev equation with and without Wigner spin
rotations.  We show and discuss results for the neutron analyzing
power $A_y$.  Sec. V contains a summary and conclusions.

\section{Poincar\'e Invariant Dynamics}

In a quantum theory the principle of special relativity requires that
the probabilities computed for equivalent experiments done in
different inertial coordinate systems are identical.  Since inertial
coordinate systems are related by Poincar\'e transformations, it
follows \cite{wigner39} that equivalent states in different inertial
coordinate systems are related by a unitary representation, $U(\Lambda
,a)$, of the Poincar\'e group.  This emphasis on the invariance of
experimental measurements in different inertial frames is different
than the covariance requirements that are historically motivated by
the way symmetries are realized in classical wave equations.

Since any representation of the Poincar\'e group can be decomposed
into a direct integral of irreducible representations, one way to
construct a Poincar\'e invariant dynamics is to build it out of
irreducible 
representations.  The transformation properties of irreducible
representations are well known and completely determined by group
theoretical considerations.  The dynamics is contained in the spectrum
of the physical mass and spin operators which determines the values
and multiplicities of the Casimir invariants that appear in this
decomposition.

Our construction begins with one-particle representations, which are
irreducible representations.  The particle's mass and spin fix the
eigenvalues of the two Casimir invariants of the Poincar\'e group.
For computations it is necessary to choose a basis for the irreducible
representation space.  This is done by choosing a maximal set of
commuting Hermitian functions of the infinitesimal Poincar\'e
generators.  In addition to the mass and spin, it is possible to find
four additional mutually commuting non-invariant functions of the
generators.  There is a second set of four operators that are
conjugate to the non-invariant commuting observables.  These operators
change the eigenvalues and determine the spectrum of the commuting
observables.  All ten generators can be expressed as functions of
these eight non-invariant operators and the two Casimir operators.
The irreducible representation space, ${\cal H}$, is the space of
square integrable functions of the eigenvalues of the four commuting
operators \cite{kei91}.

Our choice of basis for irreducible representation spaces is the
simultaneous eigenstates of the linear momentum $\mathbf{p}$ and the
3-component of the canonical spin, $j_{cz}$, which is the observable
corresponding to the spin measured in the particle's rest frame if the
particle is transformed to its rest frame with a rotationless Lorentz
transformation.  In this basis the irreducible unitary representation
of the Poincar\'e group is \cite{kei91}: 

\begin{equation}
U(\Lambda ,a) \vert (j,m) \mathbf{p} ,\mu \rangle =
\vert (j,m ) \mathbf{p}' ,\mu' \rangle e^{i p' \cdot a} 
\sqrt{\omega(p') \over \omega (p) } 
D^j_{\mu' \mu} [B^{-1} (\mathbf{p}'/m) \Lambda B(\mathbf{p}/m)]
\label{b.1}
\end{equation} 
where $p'=\Lambda p$, $\omega (p) = \sqrt{\mathbf{p} \cdot \mathbf{p}
  + m^2}$, and $B(\mathbf{p}/m)$ is the rotationless Lorentz
transformation that takes a particle of mass $m$ at rest to momentum
$\mathbf{p}$.  The quantity $R_w( \Lambda , p) := B^{-1}
(\mathbf{p}'/m) \Lambda B(\mathbf{p}/m)$ is the standard
rotationless-boost Wigner rotation.  The representation (\ref{b.1}) is
unitary for states with a $\delta (\mathbf{p}-\mathbf{p}')$
normalization in the momentum.  The important observation is that {\it
  all} mass $m>0$ spin $j$ irreducible representations of the
Poincar\'e group in the $\{ \mathbf{p},j_{cz} \}$ basis have this
form.

The two or three-nucleon Hilbert space is the tensor product of two
or three single-nucleon irreducible representation spaces; ${\cal H}
\otimes {\cal H}$ or ${\cal H} \otimes {\cal H}\otimes {\cal H} $.  On
each of these spaces
\begin{equation} 
U_0 (\Lambda ,a)=  U(\Lambda ,a) \otimes U(\Lambda ,a)
\qquad 
U_0 (\Lambda ,a) =U(\Lambda ,a) \otimes U(\Lambda ,a) \otimes U(\Lambda ,a)
\label{b.2}
\end{equation}
define kinematic representations of the Poincar\'e group.  These
representations are reducible and do not contain any dynamics.  We
build dynamical irreducible representations by adding suitable
interactions to the mass Casimir operator of non-interacting
irreducible representations.  The first step needed to introduce
interactions is then to decompose these non-interacting tensor product
representations into a direct integral of irreducible representations.
This is accomplished with Poincar\'e group Clebsch-Gordan coefficients in
our chosen $\{ \mathbf{p} , j_{cz} \}$ basis.  The 
Poincar\'e group Clebsch-Gordan
coefficients are the expansion coefficients of linear combination of 
tensor product states that transform irreducibly.  The desired
non-interacting irreducible states
are computed by (1) constructing rest eigenstates of the
two-body system, (2) decomposing them into irreducible representations
under $SU(2)$ rotations, (3) boosting the result to an arbitrary
frame. The resulting Clebsch-Gordan coefficients in this basis are
\cite{moussa}\cite{relform1}\cite{kei91}
\[
\langle  \mathbf{p}_1, \mu_1,
\mathbf{p}_2, \mu_2  \vert (j,k ) \mathbf{p} ,\mu ; l ,s \rangle
\]
\[
 =\sum_{\mu_l \mu_s \mu_1'\mu_2'}
\delta (\mathbf{p} - \mathbf{p}_1 - \mathbf{p}_2)
{\delta (k -k (\mathbf{p}_1, \mathbf{p}_2) \over k^2}
N^{-1}(p_1, p_2)   
( l, \mu_l, s, \mu_s \vert  j, \mu )
( j_1, \mu_1', j_2, \mu_2' \vert  s, \mu_s ) \times
\] 
\begin{equation}
Y_{l \mu_l }(\hat{\mathbf{k}}(\mathbf{p}_1,\mathbf{p}_2) )
D^{j_1}_{\mu_1' \mu_1}[R_w(B(\mathbf{p}/m_{120}), k_1] 
D^{j_2}_{\mu_2' \mu_2}[R_w(B(\mathbf{p}/m_{120}), k_2] =
\label{b.3}
\end{equation}
\[
 =
\int d\hat{\mathbf{k}}\sum_{\mu_l \mu_s \mu_1'\mu_2'}
\delta (\mathbf{p}_1 - \mathbf{p}_1(\mathbf{p},\mathbf{k}))
\delta (\mathbf{p}_2 - \mathbf{p}_2(\mathbf{p},\mathbf{k}))
N (p_1, p_2)   
( l, \mu_l, s, \mu_s \vert  j, \mu )
( j_1, \mu_1', j_2, \mu_2' \vert  s, \mu_s ) \times
\] 
\begin{equation}
Y_{l \mu_l }(\hat{\mathbf{k}}) 
D^{j_1}_{\mu_1' \mu_1}[R_w(B(\mathbf{p}/m_{120}), k_1] 
D^{j_2}_{\mu_2' \mu_2}[R_w(B(\mathbf{p}/m_{120}), k_2] .
\label{b.4}
\end{equation}
In these expressions 
\begin{equation}
p^{\mu} = p_1^{\mu} + p_2^{\mu} \qquad m^2_{120} = -p^2 \qquad 
k^{\mu}  = B^{-1} (\mathbf{p}/m_{120})^{\mu}{}_{\nu} {1 \over 2} (p_1-p_2)^{\nu} 
\label{b.5}
\end{equation}
\begin{equation}
N^{-2}(p_1, p_2) = {\omega (k) \omega (k) (\omega (p_1) + \omega (p_2)) \over 
\omega (p_1) \omega (p_2) (\omega (k) + \omega (k))}
\label{b.6}
\end{equation}
and the two-body invariant mass 
\begin{equation}
m_{120}:= 2\sqrt{\mathbf{k}^2 + m^2} = 2 \omega (k)
\label{b.7}
\end{equation} 
is replaced by the continuous variable $k:=\sqrt{\mathbf{k}^2}$.  The
quantum numbers $l$ and $s$ are kinematically invariant quantities
that distinguish multiple copies of representations with the same mass
($k$) and spin. For a two nucleon-system they have the same spectrum
as the orbital and spin angular momentum operators in a partial
wave representation of the nonrelativistic basis.

An irreducible representation is also obtained by changing the order
of the spin couplings in (\ref{b.3}-\ref{b.4}) where the orbital
angular momentum is first coupled to one of the spins, $j_2+l=I$, and
then the result is coupled to the second spin. $j_1+I=j$.  This
representation is constructed by making the replacements in
\[
\sum_{s,\mu_s} 
( j_1, \mu_1', j_2, \mu_2' \vert  s, \mu_s )
( l, \mu_l, s, \mu_s \vert  j, \mu )
 \to 
\]
\begin{equation}
\sum_{I,\mu_I} ( l, \mu_l', j_2, \mu_2 \vert  I, \mu_I )
( I, \mu_I, j_1, \mu_1' \vert  j, \mu )
\label{b.8}
\end{equation}
in equation (\ref{b.3}) or (\ref{b.4}).
In this representation the degeneracy parameters $(l,s)$ are replaced
by $(l,I)$.  When we construct three-body irreducible representations
by successive pairwise coupling we use the coupling (\ref{b.3}) in
the first Clebsch-Gordan coefficient and the coupling (\ref{b.8}) in
the second Clebsch-Gordan coefficient.  This allows us to identify the
quantum numbers of the relativistic irreducible basis with the quantum
numbers that we used in previous nonrelativistic calculations
(\cite{glo96,book}).

Three-particle irreducible representations for systems of
non-interacting particles can be constructed by successive pairwise
coupling of irreducible representations:
\[
\langle  
 \mathbf{p}_1, \mu_1, \mathbf{p}_2, \mu_2, \mathbf{p}_3, \mu_3 
\vert
(J, q) \mathbf{P} ,\mu ; \lambda ,I ,j_{23} ,k_{23}, l_{23} ,s_{23}
\rangle =
\]
\begin{equation}
\int d\mathbf{p}_{23}  
\sum_{\mu_{23}=-j_{23}}^{j_{23}}
\langle  
\mathbf{p}_2, \mu_2,
\mathbf{p}_3, \mu_3 
\vert 
(j_{23}, k_{23}) \mathbf{p}_{23} ,\mu_{23} ; l_{23} ,s_{23}
\rangle 
\langle 
\mathbf{p}_1, \mu_1,
\mathbf{p}_{23}, \mu_{23}
\vert 
(J, q) \mathbf{P} ,\mu ; \lambda ,I 
\rangle .
\label{b.9}
\end{equation}
For three nucleon scattering or bound state problems it is sufficient
and convenient to work in the three-body center of momentum frame.  
This simplifies the coefficients; the Wigner rotations in the second
Clebsch-Gordan coefficient in (\ref{b.9}) become the identity and the
normalization factor $N(p_{23}, p_1) \to 1$.  Both of these factors
are non-trivial in the first coefficient. 
The form of these
coefficients for the case that particle 1 is a spectator (for details
see~\cite{witrel}) and using a short hand notation, is
\begin{eqnarray}
&~&
\langle \mathbf{p}_1, \mu_1' , \mathbf{p}_2, \mu_2' ,\mathbf{p}_3, \mu_3'
\vert  (J, q) \mathbf{P}=\mathbf{0} ,\mu ; 
\lambda ,I ,j_{23} ,k_{23}, l_{23} ,s_{23} \rangle ~
=
\cr
&~&
~ \delta( \mathbf{0}-  \mathbf{q}_1 -\mathbf{q}_2
- \mathbf{q}_3 )
{1\over{ N({q}_2, {q}_3) } } \cr
&~& { \delta(q_1 -q ) \over { q^2 } }
{ \delta(k(~ \mathbf{q}_2,\mathbf{q}_3~ )-k  )
\over { k^2  } } ~ \cr
&~&
 \sum_{\mu_2 \mu_3 \mu_s }
 \sum_{\mu_l \mu_{\lambda} \mu_I }
  ( {1\over{2}}, \mu_2 ,{1\over{2}}, \mu_3 \vert s, \mu_s )
 (  l,  \mu_l, s, \mu_s, \vert j, \mu_j )
(\lambda, \mu_{\lambda}, {1\over{2}}, \mu'_1 \vert I, \mu_I )
( j, \mu_j, I, \mu_I \vert J, \mu) \cr
&~& 
 Y^{\lambda}_{\mu_{\lambda}}(\hat{\mathbf{q}}_1~)
 Y^l_{\mu_l}(\hat{\mathbf{k}}(~\mathbf{q}_2,\mathbf{q}_3 )~ )  \cr
&~&
D^{1\over{2} }_{\mu'_2 \mu_2}
[R_w(B(-\mathbf{q}_1/m_{023}),
{k}_2(~ \mathbf{q}_2,\mathbf{q}_3 )~ )] 
D^{1\over{2} }_{\mu_3' \mu_3}
[R_w (B(-\mathbf{q}_1/m_{023}) )~ ),
{k}_3(~ \mathbf{q}_2,\mathbf{q}_3 )~ )] .
\label{b.10}
\end{eqnarray}
where 
\begin{equation}
(\mathbf{q_i} ,\omega (q_i) ) = B^{-1}(\mathbf{P}/M) p_i, 
\qquad i \in \{1,2,3 \} 
\qquad
(\mathbf{k}_i ,\omega (k_i) ) = B^{-1}(-\mathbf{q}_k/m_{ij}) p_i
\label{b.11}
\end{equation}
\begin{equation}
\sum_i \mathbf{q}_i=\mathbf{0}
\qquad M = \sum_{i=1}^3  \sqrt{m^2 + \mathbf{q}_i^2} 
\label{b.12}
\end{equation}

The important property of the states 
\begin{equation}
\vert (j,k ) \mathbf{p} ,\mu ; l ,s \rangle 
\label{b.14}
\end{equation}
and 
\begin{equation}
\vert  (J, q) \mathbf{P} ,\mu ; 
\lambda ,I ,j_{23} ,k_{23}, l_{23} ,s_{23} \rangle ~
\label{b.15}
\end{equation} 
is that they transform irreducibly.  The mass and spin 
are given by 
\begin{equation}
m_{120} = 2 \sqrt{m^2 + \mathbf{k}^2} ~,~  \qquad j
\label{b.16}
\end{equation}
\begin{equation}
M =  \sqrt{4m^2 + 4\mathbf{k}^2+ \mathbf{q}^2 } +
\sqrt{m^2 + \mathbf{q}^2} ~,~ \qquad J
\label{b.17}
\end{equation}
respectively.
 
In order to use these representations to construct dynamical 
representations an interaction is added to the two- or three-body invariant 
mass operator of the form
\begin{equation}
\langle (j, k) \mathbf{p}, \mu \cdots 
\vert v \vert \cdots \mathbf{p}', \mu' (j', k')  \rangle =
\delta_{\mu \mu'} \delta_{jj'} \delta (\mathbf{p} - \mathbf{p}') 
\langle k ,\cdots 
\Vert v^j \Vert \cdots',  k' \rangle 
\label{b.18}
\end{equation}
for $N=2$ or 
\begin{equation}
\langle (J, q) \mathbf{P}, \mu \cdots 
\vert V \vert \cdots \mathbf{P}', \mu' (J', q')  \rangle =
\delta_{\mu \mu'} \delta_{JJ'} \delta (\mathbf{P} - \mathbf{P}') 
\langle k,q ,\cdots 
\Vert V^J \Vert \cdots', k',q' \rangle 
\label{b.19}
\end{equation}
for $N=3$.  Diagonalizing $m_{12} = m_{120} + v$ or $M=M_0 +V$ in
the noninteracting irreducible basis gives simultaneous eigenstates of
$m_{12},\mathbf{p}, j^2, j_z$ for $N=2$ and of $M,\mathbf{P}, J^2,
J_z$ for $N=3$.  In both the two and three-body case these eigenstates,
 $\vert  (j_{12}, \lambda_{m_{12}}) \mathbf{p}_{12} ,\mu_{12} \cdots \rangle$
 and $\vert  (J, \lambda_M ) \mathbf{P} ,\mu \cdots \rangle$,   
where $\lambda_{m_{12}}$ and $\lambda_M$  are the eigenvalues of 
 $m_{12}$ and $M$,   
are complete on the two and three-body Hilbert spaces respectively.  

The dynamical representation of the Poincar\'e group is
defined by requiring that these eigenstates transform like (\ref{b.1})
with the mass being replaced by the mass eigenvalues $\lambda_{M}$ 
or $\lambda_{m_{12}}$.
This representation is unitary and defines the dynamics
of the system.  With our choice of irreducible 
basis, $\{\mathbf{p} , j_{cz} \}$,
the resulting irreducible representations of the Poincar\'e group have
a mass independent representation of the three-dimensional Euclidean
subgroup, which Dirac \cite{dirac49} called an ``instant-form dynamics''.

For the three-nucleon case there remains the problem of how to construct
realistic $NN$-interactions.  For two-body interactions
the relation
\begin{equation}
H_{12}^2 - \mathbf{p}^2 = m_{12}^2 = 4(\mathbf{k}^2 +m^2 ) + 
4mv_{NN}= 4m ( \underbrace{\mathbf{k}^2/m + v_{NN}}_{h_{nr}=H_{nr} - {\mathbf{p}^2
\over 4m}}  
+ m ) 
\label{b.20}
\end{equation} 
implies that the square of the two-body mass operator has a simple
relation to the nonrelativistic rest Hamiltonian with a ``realistic''
NN interaction \cite{cps75}
provided one
identifies the spectrally equivalent relative momenta, $\mathbf{k}^2$.  
In the
relativistic case $k:=B^{-1} (\mathbf{p}/m_{120}) {1 \over 2} (p_1-p_2)$ 
\begin{eqnarray}
\mathbf{k} &\equiv& \mathbf{k}(\mathbf{p}_1,
\mathbf{p}_2) \cr
&=& {1\over{2}} (\mathbf{p}_1 - \mathbf{p}_2 -
\mathbf{p} {{\omega (p_2)  - \omega (p_1)} \over{\omega (p_2)  - \omega (p_1) 
+ \sqrt{(\omega (p_2)  - \omega (p_1))^2 -
\mathbf{p}^{~2}}  } }) ,
\label{b.21}
\end{eqnarray}
while in the
nonrelativistic case $k=B_g^{-1} (\mathbf{p}/2m) {1 \over 2} (p_1-p_2)$: 
\begin{equation}
\mathbf{k} = {1 \over 2}\left ( ( (\mathbf{p}_1 - {\mathbf{p} \over 2m}m) 
-(\mathbf{p}_2 - {\mathbf{p} \over 2m}m)\right ) 
= {1 \over 2} (\mathbf{p}_1-\mathbf{p}_2)  
\label{b.22}  
\end{equation} 
where
$B(\mathbf{p}/m_{120})$ is a rotationless Lorentz boost and 
$B_g (\mathbf{p}/2m)$ is the corresponding
Galilean boost.

The Kato-Birman invariance principle \cite{kato66,fcwp82,baum83} implies
that the M{\o}ller wave operators satisfy
\begin{equation}
\Omega_{\pm} (H,H_0) :=s-\lim_{t \to \pm \infty} e^{iHt}e^{-iH_0t} =
s-\lim_{t \to \pm \infty} e^{if(H)t}e^{-if(H_0)t} =
\Omega_{\pm} (f(H),f(H_0))
\label{b.23}
\end{equation}
where $f(x)$ is any piecewise differentiable function of bounded
variation with positive derivative \cite{baum83}.  The functions
$f(x)=x^2$ and $f(x)=x^{1/2}$  satisfy the conditions of
Kato-Birman theorem.  Using equation (\ref{b.23})
along with the kinematic Euclidean invariance of the
Hamiltonians $H_r$ and $H_{nr}$ gives the following relation between 
the two-body
scattering wave operators
\begin{equation}
\Omega_{\pm}(H_{r},H_{r0})=
\Omega_{\pm}(H^2_{r},H^2_{r0})=
\Omega_{\pm}(M^2_{r},M^2_{r0})=
\Omega_{\pm}(M_{r},M_{r0}).
\label{b.24}
\end{equation} 
On the other hand the identification (\ref{b.20}) along with the 
reparametrization $t \to t'= 4mt$ gives
\begin{equation}
\Omega_{\pm}(M^2_{r},M^2_{r0})= 
s-\lim_{t \to \pm \infty} e^{ih_{nr} 4mt} e^{-h_{nr0} 4mt }= 
s-\lim_{t' \to \pm \infty} e^{ih_{nr} t'} e^{-h_{nr0} t' } =
\Omega_{\pm nr} (h_{nr},h_{nr0}).
\label{b.25}
\end{equation} 
Writing 
both wave operators as direct integrals over $\mathbf{k}^2 =
m h_{0nr} = (M_0^2 - 4m^2)/4$ leads to the identifications
\begin{equation}
\Omega_{\pm} (H_{nr} , H_{0nr}) =
\Omega_{\pm} (h_{nr} , h_{0nr}) =
\int_{\oplus} \hat{\Omega}_{\pm} (\mathbf{k}^2) d\mathbf{k}^2 =
\Omega_{\pm} (H_{r} , H_{0r}) =
\Omega_{\pm} (M^2_{nr} , M^2_{0nr}) 
\label{b.26}
\end{equation}
and
\begin{equation}
S(\mathbf{k}^2) = \Omega_{r+}^{\dagger}(\mathbf{k}^2)
\Omega_{r-}(\mathbf{k}^2)=
\Omega_{nr+}^{\dagger}(\mathbf{k}^2)
\Omega_{nr-}(\mathbf{k}^2) .
\label{b.27}
\end{equation} 
The identification of the relativistic and nonrelativistic wave
operators as functions of $\mathbf{k}^2$ ensures that the relativistic
two-body model is fit to the same two-body $S$-matrix (experimental
data) as the nonrelativistic model provided the interactions are
related by (\ref{b.20}).  The identification of the wave operators
also implies the identity of the scattering wave functions as a
function of $\mathbf{k}$.  The identity of the bound state wave functions is
also due to the relation (\ref{b.20}).

In our calculations we use the interaction $v$ defined by
$v:= m_{12}-m_{120}$ which we construct \cite{kam}
from the $NN$ interaction in (\ref{b.20}) 
 by iterating  
\begin{equation}
\{ m_{120}, v \} = 4mv_{NN} - v^2  
\label{b.28}
\end{equation}
in the irreducible plane-wave basis.  Because $m_{12}$ and $m_{12}^2$
have the same eigenvectors,  the $\mathbf{k}$ dependence of the wave functions 
constructed from $m_{12}$ are also identical to the corresponding 
nonrelativistic wave functions.

The equation (\ref{b.28}) in the irreducible plane-wave basis has the form
\[
\langle k,l,s \vert v^j \vert k' ,l' ,s' \rangle =
\]
\begin{equation}
2m {\langle k,l,s \vert v_{NN}^j \vert k' ,l' ,s' \rangle \over 
\omega (k) +  \omega (k') } - 
\sum_{l''s''} \int  k^{\prime\prime~2} dk''
{\langle k,l,s \vert v^j \vert k'' ,l'' ,s'' \rangle 
\langle k'',l'',s'' \vert v^j \vert k' ,l' ,s' \rangle
\over 2 \omega (k) + 2 \omega (k')}
\label{b.29}
\end{equation}

The iteration converges quickly for realistic interactions
\cite{kam}.  While a mathematical proof of convergence of the 
iteration of (\ref{b.29}) 
is lacking,
the results of the iterations are easily tested because
the resulting  
$m_{120}+v$ must
have the same eigenfunctions as the nonrelativistic two-body
Hamiltonian.

We applied this approach using the CD~Bonn potential as the 
nonrelativistic interaction $v_{NN}(k,k')$.  In our previous studies
\cite{witrel,witbr,skibbr} we used the momentum transformation of
ref.~\cite{kam98} and in addition restricted to leading order terms
in $\mathbf{p}/\omega$ and $v/\omega$ expansion only
\begin{eqnarray}
V(\mathbf{k}, \mathbf{k}~'; \mathbf{q}~)  &=&
v(\mathbf{k}, \mathbf{k}~')~ \left(~ 1 -
{{\mathbf{q}}^{\ 2} \over{8\omega(k)\omega(k') }} ~ \right) .
\label{b.30}
\end{eqnarray}
We checked that in most cases this simple approximation leads to practically
the same results as the exact approach applied in the  present study.

Once the two-body mass operator is constructed, the three-body mass
operator for the interacting $(ij)$ pair is the well-defined
non-linear function of the two-body mass:
\begin{equation}
M_{(ij)(k)} = \sqrt{(m_{ij0} +v_{ij})^2 + \mathbf{q}_k^2} + \sqrt{m^2 + 
\mathbf{q}_k^2} .
\label{b.31}
\end{equation}
where $v_{ij}$ is embedded in the three body Hilbert space so it commutes
with $\mathbf{q}_k$.
If this is interpreted as the rest energy operator, the interacting
pair and spectator energies are additive in the rest frame.  This
implies that the $S$ matrix clusters properly in the rest frame -
while the invariance of $S$ in all frames ensures that this property
extends to all inertial coordinate systems.

Pairwise interactions in the three-body system are defined by
\[
V_{(ij)(k)} = M_{(ij)(k)}-M_0 =
\]
\begin{equation}
\sqrt{(m_{ij0} +v_{ij})^2 + \mathbf{q}^2} - \sqrt{m_{ij0}^2 + \mathbf{q}^2} .
\label{b.32}
\end{equation}

For any pair of particles these interactions commute with kinematic 
momentum and spin, and are independent on the momentum and 
magnetic quantum numbers.  This ensures that the sum of the interactions 
has the general form (\ref{b.19}).

A generalization of the method used in (\ref{b.28}-\ref{b.29}) can be used
to construct $V_{(ij)(k)}$ by iterating 
\begin{equation}
\{ \sqrt{m_{ij0}^2 + \mathbf{q}^2} , V_{(ij)(k)} \} = 
v^2 + \{ m_{ij0} , v \} - V_{(ij)(k)}^2 .
\label{b.33}
\end{equation}
Specifically 
\[
\langle k, \cdots  \vert V_{(ij)(k)}(\mathbf{q}^2)  \vert  k',  \cdots \rangle
= 
{1 \over \sqrt{m_{ij0}^2(k) + \mathbf{q}^2} + \sqrt{m_{ij0}^2(k') + \mathbf{q}^2} }
\times 
\]
\[
\left [
\langle k, \cdots \vert v^2 \vert k', \cdots  \rangle +  
2 (\omega (k) + \omega (k') ) \langle k \vert v \vert k' \rangle 
\right .
\]
\begin{equation}
\left .
-\sum_{l''s''}\int \langle k,\cdots  \vert  V_{(ij)(k)}(\mathbf{q}^2) \vert k'' ,\cdots \rangle 
k^{\prime\prime2} dk'' 
\langle k'',\cdots \vert V_{(ij)(k)}(\mathbf{q}^2) \vert k' ,\cdots \rangle 
\right ]
\label{b.34}
\end{equation}
This iteration also converges and is used to construct interactions $V_{ij}$ 
for each pair of particles.

The three-body mass (rest energy) operator is
\begin{equation}
M=M_0 + V_{12} + V_{23} + V_{31}
\label{b.35}
\end{equation}
where $M_0$ is the three-body kinematic invariant mass: 
\begin{equation}
M_{0} = \sqrt{m_{120}^2 + \mathbf{q}^2} + \sqrt{m^2 + \mathbf{q}^2} 
\label{b.36}
\end{equation}

Our relativistic Faddeev equation is based on this mass operator 
(\ref{b.35}) with
two-body interactions constructed from the CD~Bonn interaction using
equations (\ref{b.28}), (\ref{b.29}), (\ref{b.33}) and (\ref{b.34}).

Finally, just like in the nonrelativistic case, there is a natural
order of coupling of the irreducible representations for computing
each pairwise interaction.  The change of basis relating different
orders of coupling is needed for the implementation of the Faddeev
equation as an integral equation.  The required basis change only
changes the invariant degeneracy quantum numbers associated with each
order of coupling.

\begin{equation}
\langle (j m) \mathbf{P}, \mu (ab)(c) \vert  
(j' m') \mathbf{P}', \mu' (de)(f) \rangle
=
\delta (\mathbf{P} - \mathbf{P}')\delta_{jj'} \delta_{\mu \mu'} 
R^{jm} [ (ab)(c);(de)(f)] 
\label{b.37}
\end{equation}
The invariants $R^{jm} [ (ab)(c);(de)(f)]$ are Racah coefficients for
the Poincar\'e group.  They are constructed using four Poincar\'e
Clebsch-Gordan coefficients.  We compute this quantity using the
Balian-Brezin method \cite{bal}, where the variables associated with
one order of coupling are expressed in terms of the variables
associated with another order of coupling.  The invariant coefficient,
$R^{jm} [ (ab)(c);(de)(f)]$, can be computed by evaluating the
expression at zero momentum, averaging over the magnetic quantum
numbers, and evaluating the resulting expression at any kinematically
allowed set of momenta \cite{bkwp2,BBjpp}.  These Racah coefficients contain
Wigner rotations and jacobians which do not
appear in the nonrelativistic permutation operators.  Explicit expressions
are given in Appendix \ref{a2}.  

\section{Spin observables}

In a Poincar\'e invariant quantum theory or relativistic quantum field
theory the spin of a particle can be defined as the angular momentum
that is measured in the particle's rest frame.  For any non-zero
momentum, $\mathbf{p}$, different Lorentz transformations can be used
to transform a particle at rest to frame where the particle has
momentum $\mathbf{p}$.  Because the commutator of two rotationless
boost generators
\begin{equation}
[K_j,K_k] = -i \epsilon_{jkl} J_j 
\label{s.1}
\end{equation}
is a rotation generator, the spin of the particle with momentum
$\mathbf{p}$ depends on the choice of Lorentz transformation that
transforms the particle's momentum from zero to $\mathbf{p}$.  In
order to have an unambiguous definition of the spin it is necessary to
choose a standard set of $\mathbf{p}$ dependent Lorentz
boosts, $B(\mathbf{p}/m)^{\mu} {}_{\nu}$, that transform a
particle of mass $m$ at rest to momentum $\mathbf{p}$.  Then the spin
of the particle can  be unambiguously defined as the value of the spin
measured in the particles rest frame if it is transformed to the
rest frame using the standard Lorentz transformation.  With this
definition, if two spins are equal in one frame they are equal in all
frames.

The choice of standard boost is not unique because if $R(\mathbf{p}/m)$ is any
$\mathbf{p}$ dependent rotation and
\begin{equation}
B^{\prime}(\mathbf{p}/m):=B(\mathbf{p}/m) R(\mathbf{p}/m)  
\label{s.2}
\end{equation}
then both $B^{-1}(\mathbf{p}/m)^{\mu} {}_{\nu}$ and 
$B^{-1\prime}(\mathbf{p}/m)^{\mu} {}_{\nu}$ both transform 
$\mathbf{p}$ to zero.

Each choice of boost leads to a different spin operator, corresponding
to a different prescription for measuring the spin in an arbitrary
frame.  The rotation
\begin{equation}
R(\mathbf{p}/m) = B^{-1}( \mathbf{p}/m) B^{\prime}(\mathbf{p}/m)
\label{s.3}
\end{equation}
that relates different boosts is called a generalized Melosh rotation
\cite{melosh74}\cite{kei91}.

The spin operator $\mathbf{j}_x$ associated with a boost $B_x(\mathbf{p}/m)$
is defined as the following function of 
the Poincar\'e generators:
\begin{equation}
(0,\mathbf{j}_x) = B_x^{-1} (\mathbf{p}/m)^{\mu}{}_{\nu} W^{\mu} /m
\label{s.4}
\end{equation}
where $W^{\mu}= {1 \over 2} \epsilon^{\mu\alpha\beta\gamma} p_{\alpha}
M_{\beta \gamma}$ is the Pauli-Lubanski vector, $M_{\beta \gamma}$ 
is the relativistic angular momentum tensor,
and $B_x (\mathbf{p}/m)$
is the boost matrix with the parameters $\mathbf{p}/m$ and $m$ replaced by 
the mass and momentum operators.

While this quantity has the appearance of a four vector, it is not
because of the operator dependence of the arguments of the boost.
Instead, under Lorentz transformations the spin transforms like
\begin{equation}
U^{\dagger} (\Lambda,0) \mathbf{j}_x U(\Lambda ,0)=
R_{wx} (\Lambda ,p)  \mathbf{j}_x   
\label{s.5}
\end{equation}
where 
\begin{equation}
R_{wx} (\Lambda ,p) :=
B_x ^{-1} (\mathbf{\Lambda p}/m)\Lambda B_x(\mathbf{p}/m)   
\label{s.6}
\end{equation}
is the Wigner rotation associated with the boost $B_x(\mathbf{p}/m)$.
It is a consequence of the Poincar\'e commutation relations that the
components of any of these spin observables satisfy $SU(2)$
commutation relations
\begin{equation}
[j_{xl},j_{xm} ]_{-} = i \epsilon_{lmn} j_{xn} 
\label{s.7}
\end{equation}
for any ``$x$''.  The operator $\mathbf{j}^2$ is independent of the
choice of boost because the generalized Melosh rotations leave the
scalar product of two vectors unchanged.  The spin defined with the
textbook rotationless or canonical boost is called the canonical spin.

It is natural to ask how these different types of spins are measured
in the laboratory.  Spins of isolated elementary or composite
particles are measured in the laboratory through their response to
classical electromagnetic fields.  In the one-photon exchange
approximation the photon couples to matrix elements of a covariant
current operator.  Imposing Poincar\'e covariance, current
conservation, and discrete symmetries allows one to express all
current matrix elements in terms of an independent set of matrix
elements, which have a 1-1 correspondence with invariant form factors.
All conventional form factors can be expressed in terms of Breit frame
matrix elements with canonical spin and a quantization axis parallel
to the Breit frame momentum transfer.

In the $SL(2,C)$ representation canonical boosts are represented by 
positive Hermitian matrices.  They have the general
form
\begin{equation}
B (\mathbf{p}/m) = exp (\bsig \cdot \brho /2)
\label{s.8}
\end{equation}
where $\brho$ is the rapidity of the Lorentz transformation
and $\bsig$ are the Pauli matrices. 
In this paper all of our spins are canonical spins. The 
$SO(1,3)$ representation of canonical boosts are
the standard rotationless boosts. 

Given a definition of the form factors in terms of
independent current matrix elements in a given basis, it is also
possible to express them in terms of current matrix elements in any
other standard frame using any other basis~\cite{chung1988}.  For example the
expression in terms of Breit frame canonical spin matrix elements, can
be replaced by a different independent set of laboratory frame
helicity spin matrix elements.  In quantum field theory 
the choice of boost is built into conventions used to define 
the Dirac spinors.  The relation of the invariant form factors to 
current matrix elements with different choices of spin determines
the relationship between different spin observables and experiment. 

The spin degrees of freedom of the asymptotic incoming or outgoing
particles are most conveniently expressed in terms of traces of
density matrices, which is a reflection of the fact that 
realistic initial and/or final states are generally not pure states.  
Scattering spin observables in cross sections are formally defined by
\cite{glo96} 
\begin{equation}
\langle O \rangle = {\mbox{Tr} (S_f T S_i T^{\dagger})
\over \mbox{Tr} ( T T^{\dagger})} 
\label{s.9}
\end{equation}
where $T$ is the invariant scattering amplitude for the 
reaction under consideration.  The connection between 
the invariant scattering amplitude defined in the particle
data book \cite{pdb} and the transition amplitudes constructed by 
solving our formulation of the relativistic Faddeev equations is 
given in \cite{tlce}.  

The quantities $S$ have the form 
\begin{equation}
S = \sum s_a S_a 
\label{s.10}
\end{equation}
where the index runs over all $N_i$ or $N_f$ initial or final sets of
magnetic quantum numbers, $S_a$ are a basis for $N_i\times N_i$ or
$N_f\times N_f$ matrices that are orthonormal with respect to the
trace norm and $s_a$ are constant coefficients \cite{glo96} .

If  the initial  and  final  asymptotic states  are  represented in  a
canonical spin basis, then the magnetic quantum numbers that appear in
the  invariant amplitudes  $T$ transform  with Wigner  rotations under
Lorentz transformations.  The result is that the spin observable $\langle O
\rangle$ will not be invariant unless the matrices $S_a$ or coefficients
$s_a$ are defined to transform in a manner that leaves the observable 
invariant.

Any spin observable can be made Lorentz invariant,  by defining 
the invariant observable as its value in a given frame if it is transformed 
to the frame with a specific Lorentz boost.  This can be used
to get an invariant definition of the vector or tensor polarizations.

In this paper invariant spin observables are defined to be the values
of the observable in the laboratory frame (rest frame of the target).
To evaluate the corresponding spin observable it is only necessary 
to evaluate the expression
\begin{equation}
\langle O \rangle = {\mbox{Tr} (S_f  T  S_i  T^{\dagger} )
\over  \mbox{Tr} (T T^{\dagger} ) }
\label{s.11}
\end{equation}
for the values of the invariant amplitudes with laboratory 
kinematics.

This observable is equal its value evaluated in other frames 
using the formula 
\begin{equation}
\langle O \rangle = {Tr (S_f D^{\dagger} M D' S_i D^{\prime\dagger} M^{\dagger}
D ) \over \mbox{Tr} (M M^{\dagger} ) }
\label{s.12}
\end{equation}
where the invariant amplitudes are evaluated in the other frame and
$D$ and $D'$ are products of Wigner $D$-functions of the
Wigner rotations associated with the
boost from the lab frame to the other frame.

Our specific interest in this paper is the observable $A_y$
with polarized  incoming nucleon.  The convention used to define $A_y$ 
is the Madison convention, where the laboratory frame 
scattering plane is in the $xz$ plane.  The observable $A_y$, 
is defined as 
\begin{equation}
\langle A_y \rangle = {\mbox{Tr} (T (\sigma_y \times I_d)  T^{\dagger})
\over \mbox{Tr} ( T T^{\dagger})} 
\label{s.13}
\end{equation}
Because  the Wigner rotation for 
canonical boost along the direction of a particle's momentum 
is the identity,  $\sigma_y$ 
 is unchanged so  $A_y$ can also be evaluated in the c.m. frame without making 
any compensating Wigner rotations.  

\section{Faddeev Equation}

The nucleon-deuteron scattering with neutron and protons interacting
through a NN interaction $v_{NN}$ alone is described in terms of a
breakup operator $T$ satisfying the Faddeev-type integral
equation~\cite{wit88,glo96}

\begin{eqnarray}
T\vert \phi \rangle  &=& t P \vert \phi \rangle + t P G_0 T \vert \phi 
\rangle .
\label{eq1a}
\end{eqnarray}

The two-nucleon (2N) $t$-matrix $t$ results from solving the
Lippmann-Schwinger equation with the interaction $v_{NN}$.  The
permutation operator $P=P_{12}P_{23} + P_{13}P_{23}$ is given in terms
of the transposition $P_{ij}$ which interchanges nucleons i and j. The
incoming state $ \vert \phi \rangle = \vert \mathbf{q}_0 \rangle \vert
\phi_d \rangle $ describes the free nucleon-deuteron motion with
relative momentum $\mathbf{q}_0$ and the deuteron state vector $\vert
\phi_d \rangle$.  Finally $G_0$ is resolvent of the three-body center
of mass kinetic energy.

The  elastic nd scattering transition operator U
is  given in terms of $T$ by~\cite{wit88,glo96}
\begin{eqnarray}
U  &=& P G_0^{-1} + P T  .
\label{eq1c}
\end{eqnarray}

This is our standard nonrelativistic formulation, which is equivalent
to the nonrelativistic 3N Schr\"odinger equation plus boundary
conditions.  The formal structure of these equations in the
relativistic case remains the same but the ingredients change. As
explained in ~\cite{relform} the relativistic 3N rest Hamiltonian
(mass operator) has the same form as the nonrelativistic one, only
the momentum dependence of the kinetic energy changes and the relation
of the pair interactions in the three-body problem to the pair
interactions in the two-body problem changes. Consequently all the
formal steps leading to Eqs.(\ref{eq1a}) and (\ref{eq1c}) remain the
same.

The free relativistic invariant mass of three identical nucleons 
  in their c.m. system has the form (\ref{b.36}) 
while the free two-body mass operator has the form (\ref{b.7}).

As introduced in ~\cite{relform1} the pair forces in the relativistic
3N $2+1$ mass operator are given by (\ref{b.32}) where $V=
V(\mathbf{q}^2)$ reduces to the interaction $v$ for $\mathbf{q} = 0$.
 
The transition matrix that appears in the kernel of the Faddeev
equation (\ref{eq1a}) is obtained by solving the ``Lippmann-Schwinger
equation'', which must be solved as a function of $\mathbf{q}^2$
\begin{eqnarray}
 t( \mathbf{k}, \mathbf{k}~' ;  \mathbf{q}^2~) &=&
 V( \mathbf{k}, \mathbf{k}~' ;  \mathbf{q}^2~) +
\int d^3k'' {
{V( \mathbf{k}, \mathbf{k}~'' ; \mathbf{q}^2~)
 t( \mathbf{k}~'', \mathbf{k}~' ;  \mathbf{q}^2~) }
\over{ \sqrt{ {( 2\omega( {\mathbf{k}}^{\ '} )^{\ 2} + {\mathbf{q}}^{\ 2}} }
-  \sqrt{ {(2\omega(  {\mathbf{k}}^{\ ''})^{\ 2}
+ {\mathbf{q}}^{\ 2}} } + i\epsilon  } } .
\label{eq2a}
\end{eqnarray}
The input two-body interactions are computed by solving equation
(\ref{b.33}) and (\ref{b.34}).

The new relativistic ingredients in Eqs.(\ref{eq1a})
and (\ref{eq1c}) will therefore
be the
$t$-operator (\ref{eq2a}) (expressed in partial waves) 
and the resolvent of the 3N invariant mass
\begin{eqnarray}
G_0 &=& {{1}\over{E + i\epsilon - M_0}} ,
\label{eq2c}
\end{eqnarray}
where $M_0$ is given by Eq.~(\ref{b.36}) and $E$ is the total 3N c.m.
energy expressed in terms of the initial
neutron momentum $\mathbf{q}_0$ relative to the deuteron
\begin{eqnarray}
E &=& \sqrt{(M_d)^2 + {\mathbf{q}}_0^{\ 2} } + \sqrt{m^2 + 
{\mathbf{q}}_0^{\ 2}} ,
\label{eq2d}
\end{eqnarray}
and $M_d$ is the deuteron rest mass.

Currently Eq.(\ref{eq1a}) in its nonrelativistic form is numerically
solved for any NN interaction using a momentum space partial-wave
decomposition.  Details are presented in ref.~\cite{wit88}. This turns
Eq.(\ref{eq1a}) into a coupled set of two-dimensional integral
equations.  As shown in~\cite{witrel}, in the relativistic case we can
keep the same formal structure, though the permutation operators are
replaced by the corresponding Racah coefficients (\ref{b.37}) for the
Poincar\'e group.  These coefficients include both jacobians and
Wigner rotations that do not appear in the nonrelativistic
permutation operators~\cite{glo96,book}.  
These coefficients are computed in the
Appendix \ref{a2}, using methods that we have applied to compute the
nonrelativistic permutation operators.

In the nonrelativistic case the partial wave projected momentum space basis is
\begin{eqnarray}
 &~& \vert p q (ls)j
(\lambda{1\over{2}})IJ(t{1\over{2}})T \rangle ,
\label{eq2e}
\end{eqnarray}
where p and q are the magnitudes of standard Jacobi momenta (see
\cite{glo96,book}), obtained by transforming single particle momenta to the
rest frame of a two or three-body system using Galilean boosts, and
$(ls)j$ two-body quantum numbers with obvious meaning, $(\lambda
1/2)I$ refer to the third nucleon (described by the momentum q), $J$
is the total 3N angular momentum and the rest are isospin quantum
numbers.  In the relativistic case this basis is replaced by the
irreducible plane wave states defined in (\ref{b.10}).

The basis states (\ref{b.10}) are used for
the evaluation of the partial wave representation of
the permutation operator P with Wigner rotations of spin states for nucleons
$2$ and $3$ included.  In the relativistic case we adopt 
the following short-hand notation for the
irreducible three-body states, which also includes isospin quantum
numbers coupled in the same order:
\begin{equation}
\vert k, q ,\alpha \rangle :=
\vert k q (ls)j (\lambda ,{1\over{2}}) IJ(t{1\over {2}} )T \rangle =
\vert (J, q) \mathbf{P} ,\mu ; \lambda ,I ,j_{23} ,k_{23}, l_{23} ,s_{23}
\rangle \vert  (t{1\over {2}} ) T \rangle
\label{b.13}
\end{equation}
Equipped with that, projecting  Eq.(\ref{eq1a}) onto the basis states
$\vert k, q ,\alpha \rangle$
 one encounters like in the nonrelativistic notation ~\cite{book}
\begin{eqnarray}
_1\langle k q \alpha \vert P \vert k' q' \alpha' \rangle_1
&=&  _1\langle k q \alpha \vert k' q' \alpha' \rangle_2 +
  _1\langle k q \alpha \vert k' q' \alpha' \rangle_3
 =  2~  _1\langle k q \alpha \vert k' q' \alpha' \rangle_2 .
\label{eqm6}
\end{eqnarray}
This is evaluated by inserting the complete basis of states $\vert
\mathbf{p}_1, \mu_1 ,\mathbf{p}_2, \mu_2, \mathbf{p}_3 ,\mu_3 >$ and
using (\ref{b.10}).   It can
be expressed in a form which resembles closely the one appearing in
the nonrelativistic regime ~\cite{book,glo96}
\begin{eqnarray}
_1\langle k ~ q ~ \alpha \vert ~ P ~ \vert k' ~ q' ~ \alpha' \rangle_1 &=&
\int_{-1}^{1} dx { {\delta(k-\pi_1)} \over { k^{2} }  } ~
{ {\delta(k'-\pi_2)} \over { k'^{2} }  } ~  \cr
&~&
{1\over {N_1(q, q',x) } } ~ {1\over {N_2(q, q',x) } } ~
G_{\alpha \alpha'}^{BB} (q, q', x) ,
\label{eqm7}
\end{eqnarray}
where all ingredients are given in the Appendix \ref{a2}.

Due to short-range nature of the NN interaction it can be considered
negligible beyond a certain value $j_{max}$ of the total angular
momentum in the two-nucleon subsystem. Generally with increasing
energy $j_{max}$ will also increase. For $j > j_{max}$ we put the
t-matrix to be zero, which yields a finite number of coupled channels
for each total angular momentum J and total parity
$\pi=(-)^{l+\lambda}$ of the 3N system.  To achieve converged results
at our energies we used all partial wave states with total angular
momenta of the 2N subsystem up to $j_{max}=5$ and took into account
all total angular momenta of the 3N system up to $J=25/2$. This leads
to a system of up to 143 coupled integral equations in two continuous
variables for a given $J$ and parity.

\section{Results and discussion}

The subject of the present study is to investigate the influence of
relativity on the nd elastic scattering nucleon analyzing power $A_y$
at low energies.  We define the invariant observable $A_y$ to be the
value measured in the laboratory frame (target at rest).

To this aim
we solved Faddeev equations at a number of the incoming neutron lab
energies $E_n =5$~MeV, $8.5$~MeV, and $13$~MeV. To check the energy
dependence of the effect we added two additional energies $E_n
=35$~MeV and $65$~MeV. In order to see the importance of specific
relativistic features we solved the equation in the relativistic case with
and without Wigner rotations. This allowed us to see which effects,
dynamical corrections, which are induced by the momentum dependence of 
the two-body force together with kinematical relativistic effects
coming from the use of the Poincar\'e Jacobi variables, 
 or Wigner rotations, dominate for $A_y$.

\begin{figure}
\includegraphics[scale=0.6]{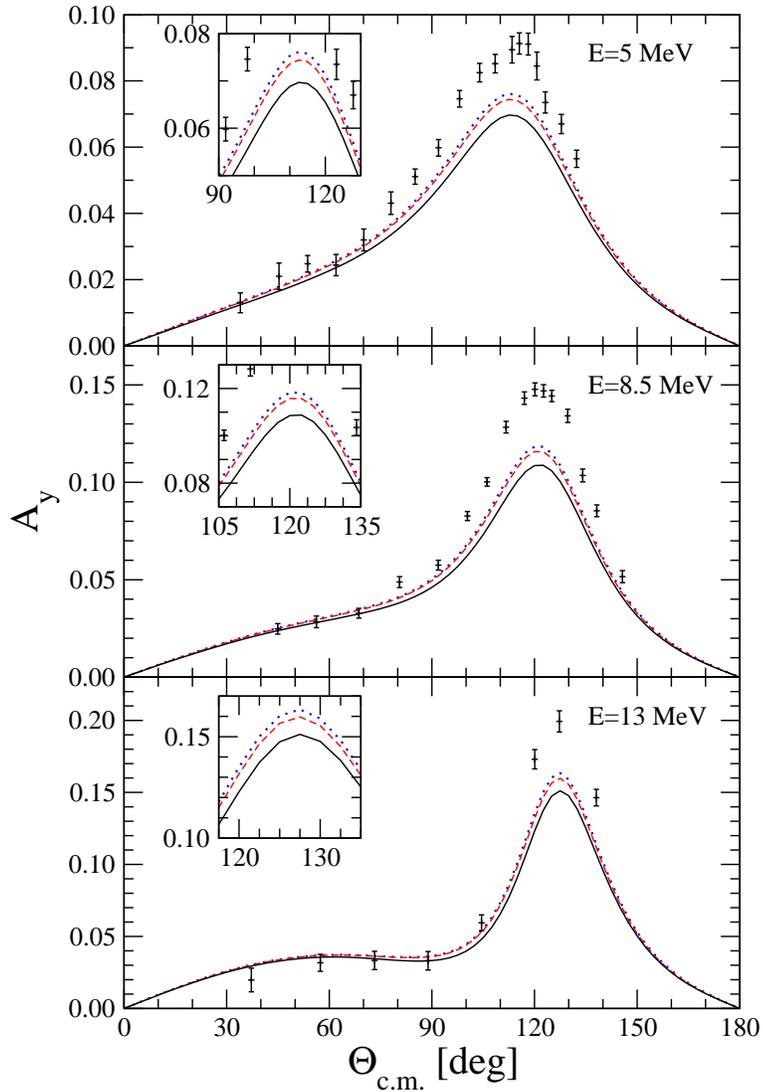}% Here is how to import EPS art
\caption{(color online)
The nucleon analyzing power $A_y$ for $nd$ elastic scattering at
various lab energies $E_n^{lab}$ of the incoming neutron.  The dotted
line is the result of the nonrelativistic Faddeev calculation with
the CD~Bonn potential. The relativistic predictions without and with
Wigner spin rotations are shown by the dashed and solid lines,
respectively.  The nd data at $5$~MeV and $8.5$~MeV are from ref.~\cite{tunl1}
and at $13$~MeV are from ref.~\cite{erl}.
}
\label{fig1}
\end{figure}

\begin{figure}
\includegraphics[scale=0.6]{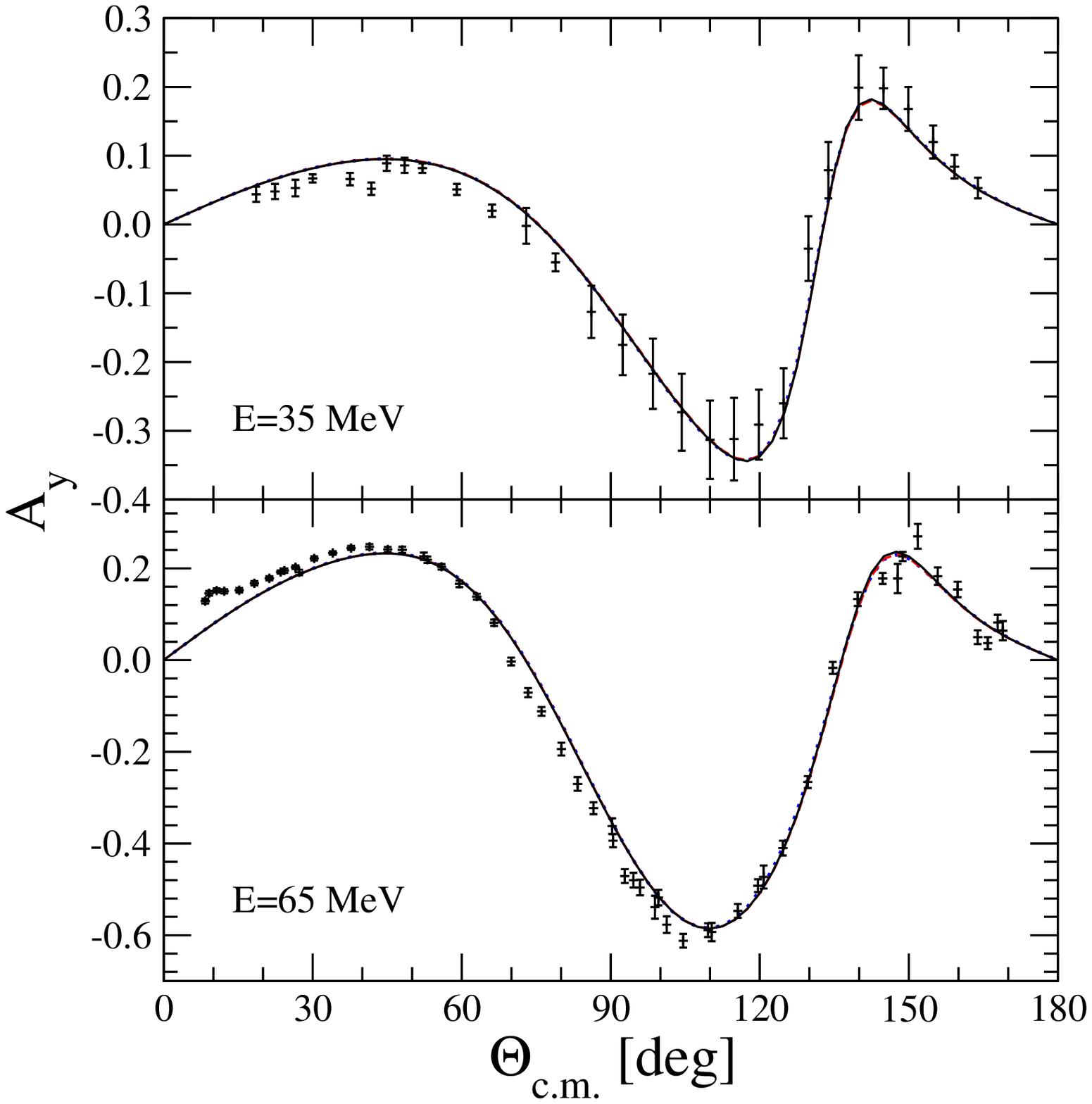}% Here is how to import EPS art
\caption{(color online) 
The nucleon analyzing power $A_y$ for $nd$ elastic scattering at
$E_n^{lab}=35$~MeV and $65$~MeV.  For description of lines see
Fig.\ref{fig1}. All theoretical predictions are practically
overlapping.  The pd data at $35$~MeV are from ref.~\cite{bunker} and
at $65$~MeV from \cite{shi95}.
}
\label{fig2}
\end{figure}

Figs.\ref{fig1} and \ref{fig2} illustrate the results.
   When only
dynamical effects are taken into account (Wigner rotations neglected) 
then at low energies of the
incoming neutron the relativistic and nonrelativistic predictions are
practically the same with the exception of the angular region close to
the maximum of $A_y$ where the relativistic prediction is $\approx
2\%$ below the nonrelativistic one (see Fig.\ref{fig1}). This small
effect disappears at higher energies (see Fig.\ref{fig2}). Including
Wigner rotations lowers significantly the values of $A_y$ in a large
region of angles around the maximum. The changes in the maximum are up
to $\approx 10\%$. Again, when the energy of the incoming neutron
increases nonrelativistic and relativistic predictions are
practically identical.

The large changes of $A_y$ occur in a region of energies where this
observable is extremely sensitive to changes in $^3P_0$, $^3P_1$, and
$^3P_2-^3F_2$ NN force components \cite{tor}. At energies where this
sensitivity dies out also the relativistic effects for $A_y$ become
negligible. This allows to conclude that the large effects seen for
$A_y$ at low energies are due to amplification of changes of the
$^3P_j$ contributions due to relativity by a large sensitivity of
$A_y$ to $P$-waves.

In a recent study \cite{schwenk} the changes of $A_y$ due to
relativity by $10\%$ at $E_n^{lab}=3$~MeV have been reported. Very
probably the opposite sign of the effect found in that study can be
attributed to the impulse approximation used when calculating $A_y$.

%\subsection{Results for the elastic Nd scattering}

\section{Summary and outlook}

We numerically solved the 3N Faddeev equation for nd scattering
including relativistic kinematics, dynamical relativistic effects and
Wigner rotations at the neutron lab energies $E_n^{lab} = 5$, $8.5$,
$13$, $35$, and $65$~MeV.  As dynamical input we took the
nonrelativistic NN potential CD~Bonn and generated in the 2N c.m.
system an exactly on-shell equivalent relativistic interaction $v$, by
solving numerically nonlinear quadratic equation relating matrix
elements of the nonrelativistic and relativistic potentials.  We
checked that the approximate procedure using an analytical scale
transformation of momenta applied in our previous studies provides
practically the same results as the present exact approach. 
 In addition a similar nonlinear equation (Eq.(\ref{b.33}) ) 
was used to generate the 
 momentum dependent two-body interaction embedded in the $3$-particle 
Hilbert space.

We found that at low energies the effects of Wigner rotations are most
important for the analyzing power.  
 They lower the maximum of $A_y$ by
up to $\approx 10\%$.  The dynamical relativistic effects are of minor
importance for $A_y$ and provide small changes of $A_y$ in a region
close to its maximum. They lower $A_y$ by only $\approx 2\%$. The
relativistic effects disappear at higher energies.

Wigner rotations  are negligible for
the cross section and all other spin observables in elastic Nd scattering 
 with exception of 
four low energy spin correlations $C_{z,yz}$, $C_{x,xy}$,  
$C_{y,xx-yy}$, $C_{y,yy}$ and 
four low energy spin transfer coefficients: 
from deuteron to deuteron  $K_{z}^{y'z'}$ and 
  $K_{x}^{x'y'}$, and from deuteron to neutron  
 $K_{yz}^{z'}$ 
and $K_{xy}^{x'}$. 
Very probably this can be traced back 
 as in the case of $A_y$ to their  sensitivity 
to  $^3P$-waves.

These results shed new light on the low-energy analyzing power
puzzle. It is known that the existing discrepancies between $A_y$ data
and theoretical predictions based on NN potentials only cannot be
removed when current three-nucleon force, mostly of $2\pi$-exchange
character \cite{TM,uIX} are included in the nuclear Hamiltonian. This
indicated that additional 3N forces should be added to the
$2\pi$-exchange type forces. Such forces provided by $\chi$PT in NNLO
and NNNLO orders are expected to provide the solution for the $A_y$
puzzle~\cite{epel2002,epelnew}.  It seems that 
in view of the present result they
must increase the maximum of $A_y$ stronger than expected up to now. 
However, $A_y$ is a very sensitive observable and our approach, using
the irreducible $\{ \mathbf{p},j_{cz}\}$ basis, is only one of many possible
basis choices that lead to Poincar\'e invariant dynamical theories
that are two-body scattering equivalent, which may give different
three-body predictions.

\section*{Acknowledgments}

This work was  partially supported 
by the Helmholtz Association through 
funds provided to the virtual institute ``Spin and strong QCD''(VH-VI-231).  
It was also performed in part under
the auspices of the U.~S.  Department of Energy, Office of Nuclear
Physics, contract No. DE-FG02-86ER40286 with the University of Iowa. 
The numerical calculations were performed on the IBM Regatta p690+ of
the NIC in J\"ulich, Germany.  

\appendix 

\section{spinors}
\label{a1}

In our calculations $SL(2,\mathbb{C})$ matrices are used to represent Wigner
rotations and Lorentz transformations.  In this way spin algebra is
reduced to working with $2 \times 2$ complex matrices, and there is no
need for representations in terms of Euler angles.  The relevant
relations are give below.

The coordinates of a four-vector 
$p^{\mu}$ can be labeled by the $2 \times 2$
Hermitian matrix $P$:
\begin{equation}
P := p^{\mu} \sigma_{\mu} =
\left (
\begin{array}{cc}
p^0+p^3 & p^1-ip^2 \\
p^1+ip^2 & p^0 - p^3
\end{array} 
\right ), 
\label{a2.1}
\end{equation}
where $\sigma_{\mu}$ are the identity and the three Pauli matrices. 
The components of $p^{\mu}$ can be extracted from the matrix $P$ using: 
\begin{equation}
p^{\mu} = {1 \over 2} \mbox{Tr} (\sigma_{\mu} P) = 
{1 \over 2} \mbox{Tr} (P \sigma_{\mu} ) . 
\label{a2.2}
\end{equation}

Because
\begin{equation}
\det (P) = (p^0)^2 - (\vec{p}\,)^2 = - \eta_{\mu \nu} p^{\mu}p^{\nu} = m^2  
\label{a2.3}
\end{equation}
and 
\begin{equation}
P = P^{\dagger} 
\label{a2.4}
\end{equation}
for real $p^{\mu}$, it follows that any linear transformation that
preserves the Hermiticity and determinant of $P$ is a real Lorentz
transformation.  It is easy to show that 
if $A$ is a complex $2\times 2$ matrix with $\det (A)=1$ then the 
transformation
\begin{equation}
P \to P' = APA^{\dagger}
\label{a2.5}
\end{equation}
has both of these properties: 
\begin{equation}
\det (P')= \det (P) \qquad  \mbox{and} \qquad 
P=P^{\dagger} \to P'=P^{\prime\dagger}.
\label{a2.6}
\end{equation}

The most general $2 \times 2$ matrix with determinant 1 can be written as
\begin{equation}
A = \pm e^{{i\over 2}\bsig \cdot \mathbf{z} }
\label{a2.7}
\end{equation}
where $\mathbf{z} = \btheta - i \brho$.  If $\brho=\mathbf{0}$
then $A= U(\btheta )$ is unitary and 
corresponds to an $SU(2)$ rotation 
through an angle $\vert \btheta \vert$ about the $\hat{\btheta}$ axis.
If $\btheta=\mathbf{0}$ then $A$ is a positive Hermitian matrix that
corresponds to a rotationless (canonical) Lorentz boost with 
rapidity $\vert \brho \vert$ in the direction $\hat{\brho}$.

The rotation $U(\btheta )$ is given by
\begin{equation}
A \to U(\btheta) = \sigma_0 \cos ({\theta \over 2}) + i \bsig \cdot
\hat{\btheta} \sin ({\theta \over 2}).
\label{a2.8}
\end{equation}
The axis of rotation can be extracted using
\begin{equation}
\hat{\btheta} = -i {\mbox{Tr} (\bsig U(\btheta) ) 
\over \vert  \mbox{Tr} (\bsig U(\btheta) )\vert} 
\label{a2.9}
\end{equation}
and the angle of rotation can be extracted from 
\begin{equation}
\theta = 2 \tan^{-1}\left ({ \mbox{Tr} (-i\bsig\cdot \hat{\btheta}  U(\btheta))
\over \mbox{Tr} ( U(\btheta))}\right ) .
\label{a2.10}
\end{equation}

The rotationless boost that transforms a particle of mass $m$ at rest
to total momentum $\mathbf{p}$ can be labeled by the final four velocity 
$\mathbf{q} := \mathbf{p}/m$:
\begin{equation}
A\to  B(\mathbf{q}) =
\sigma_0 \cosh ({\rho \over 2}) +
\bsig \cdot \hat{\mathbf{p}} \sinh ({\rho \over 2})  
\label{a2.11}
\end{equation}
where $\brho = \hat{\mathbf{p}}\rho$ is the rapidity of the 
Lorentz boost which is related to $\mathbf{q}$ by  
\begin{equation}
\cosh ({\rho \over 2}) =
\sqrt{{q^0+1 \over 2}} =
\sqrt{{p^0+m \over 2m}}
\label{a2.12}
\end{equation}
and 
\begin{equation}
\sinh ({\rho \over 2})  = {\vert \mathbf{q} \vert \over \sqrt{2(q^0+1)}} =
{\vert \mathbf{p} \vert  \over \sqrt{2m(p^0+m)}} .
\label{a2.113}
\end{equation}
$B(\mathbf{q})$  satisfies
\begin{equation}
B(\mathbf{q}) \left (
\begin{array}{cc}
m & 0 \\
0 & m
\end{array} 
\right ) 
B^{\dagger} (\mathbf{q}) =
\left (
\begin{array}{cc}
p^0 +p^3 & p^1-ip^2 \\
p^1+ip^2 & p^0 - p^3
\end{array} 
\right ) =P 
\label{a2.14}
\end{equation}
where
\begin{equation}
p^0 = \omega (p) = \sqrt{m^2 + \mathbf{p}\cdot \mathbf{p} } .
\label{a2.15}
\end{equation}
The rotationless boost above is called the canonical boost.
The inverse transformation is obtained by reversing the sign of
$\mathbf{q}$ or $\mathbf{p}$.

The Wigner $D$ functions are homogeneous polynomials 
of degree $2j$ in the coefficients of the $SL(2,\mathbb{C})$
matrices $A_{ij}$:
\begin{equation}
D^j_{\mu \nu} (A) :=
\sum_{\alpha=0}^{2j} {[(j+\mu)! (j-\mu)! (j+\nu)! (j-\nu)!]^{1/2}  \over
(j+\mu-\alpha)!\alpha! (\alpha - \mu + \nu)! (j -\nu -\alpha)!}
A_{11}^{j+\mu-\alpha}   
A_{12}^{\alpha}   
A_{21}^{\alpha -\mu+\nu}   
A_{22}^{j-\nu-\alpha}   
\label{a2.16} 
\end{equation}
These are representations of both $SL(2,\mathbb{C})$ and $SU(2)$. The $j=1/2$
representation is just the matrix $A$. 

%\appendix

\section{Permutation operator}
\label{a2}

Using Eq.~(\ref{b.10}) twice for the bra state
$ _1 \langle  k, q, \alpha \vert $ and
the ket state
$ \vert k^\prime, q^\prime, \alpha^\prime  \rangle_2 $
one gets for the matrix element of the permutation
operator in our partial wave basis:

\begin{eqnarray}
&~& _1\langle k ~, q ~, \alpha \vert ~ P ~ \vert k'~, q,' ~ \alpha' 
\rangle_1 =
2~ _1\langle k ~, q ~, \alpha ~ \vert ~ k', ~ q', ~ \alpha' \rangle_2 
= \cr
&~&
2~ \sum_{m_1 m_2 m_3 }
\sum_{\mu_2 \mu_3 \mu_s }
\sum_{\mu_l  \mu_{\lambda} \mu_I \mu}
\sum_{\mu_2' \mu_3' \mu_{s'} }
\sum_{\mu_{l'}  \mu_{\lambda'} \mu_{I'} \mu'} ~ \cr
&~&
(\lambda \mu_{\lambda} {1\over{2}}, m_1, \vert I, \mu_I ) ~
( j, \mu, I ,\mu_I \vert J, M)  ~
( {1\over{2}}, \mu_2, {1\over{2}}, \mu_3 \vert s, \mu_s ) ~
(  l,  \mu_l, s, \mu_s \vert j, \mu ) \cr
&~&
(\lambda' ,\mu_{\lambda'}, {1\over{2}}, m_2, \vert I', \mu_{I'} ) ~
( j', \mu', I', \mu_{I'} \vert J, M)  ~
( {1\over{2}}, \mu_2', {1\over{2}}, \mu_3' \vert s', \mu_{s'} ) ~
(  l',  \mu_{l'}, s', \mu_{s'} \vert j', \mu' )  \cr
&~&
 \int d\hat{\mathbf{q}} ~d\hat{\mathbf{q}}'~
{1\over{ N(\mathbf{q}~', -\mathbf{q} - \mathbf{q}~') } }  ~
{1\over{ N( -\mathbf{q} - \mathbf{q}~', \mathbf{q}~ ) } }
\cr
&~&
{ \delta(~k -
\vert \mathbf{k}(~ \mathbf{q}~', -\mathbf{q} -
\mathbf{q}~'~ ) \vert ~ )
\over { \mathbf{k}^2 } } ~
{ \delta( ~ k' -
\vert \mathbf{k}(~ -\mathbf{q} -
\mathbf{q}~', \mathbf{q} ~ ) \vert ~ )
\over { {\mathbf{k}'}^2 } } ~ \cr
&~&
Y^{\lambda~*}_{\mu_{\lambda}}(\hat{\mathbf{q}}) ~
 Y^{l~*}_{\mu_l}(\hat{\mathbf{k}}(~ \mathbf{q}~', -\mathbf{q}
-  \mathbf{q}~' )~ ) ~
Y^{\lambda'}_{\mu_{\lambda'}}(\hat{\mathbf{q}}~') ~
 Y^{l'}_{\mu_{l'}}(\hat{\mathbf{k}}(~ -\mathbf{q}
-\mathbf{q}~', \mathbf{q}~ )~ ) ~  \cr
&~&
D^{{1\over{2}}~*}_{m_2 \mu_2}
[R_w(B( -\mathbf{q}/2\omega_m(\mathbf{k}) ),
(\mathbf{k}(~ \mathbf{q}~', -\mathbf{q}
-\mathbf{q}~' )~,\omega_m(\mathbf{k}) )] ~ \cr
&~&   
D^{{1\over{2}}~*}_{m_3 \mu_3}
[R_w(B( - \mathbf{q}/2\omega_m(\mathbf{k}) ),
(- \mathbf{k}(~ \mathbf{q}~', -\mathbf{q}
- \mathbf{q}~'~)~,\omega_m(\mathbf{k})  )] ~  \cr
&~&
D^{{1\over{2}}}_{m_3 \mu_2'}
[R_w(B( -\mathbf{q}~'~/2\omega_m(\mathbf{k}) ), 
(\mathbf{k}(~ -\mathbf{q}
-\mathbf{q}~', \mathbf{q}~ )~, \omega_m(\mathbf{k}) )] ~  \cr
&~&
D^{{1\over{2}}}_{m_1 \mu_3'}
[R_w(B( -\mathbf{q}~'~/2\omega_m(\mathbf{k}) ),
(-\mathbf{k}(~ -\mathbf{q}
-\mathbf{q}~', \mathbf{q}~ )~, \omega_m(\mathbf{k}) )] ~ \cr
&~&
 _1< (~ (t{1\over{2}}) T ~ \vert ~ (t'{1\over{2}}) T ~ >_2 ,
\label{eq17}
\end{eqnarray}
with
\begin{eqnarray}
&~& \mathbf{k}(~\mathbf{q}~' , -\mathbf{q}
-\mathbf{q}~')~ \equiv ~ \mathbf{q}~' +
{1\over{2}} \mathbf{q} (1 + y_1(q, q',x))  ~ \cr
&~&
\mathbf{k}(~ -\mathbf{q}
-\mathbf{q}~', \mathbf{q}~ )~ \equiv ~ -\mathbf{q} -
{1\over{2}} \mathbf{q}~' (1 + y_2(q, q',x)) .
\label{eq18}
\end{eqnarray}
In Eq.(\ref{eq18}) occur
\begin{eqnarray}
y_1(q, q',x) =
\frac{  E_{ {\mathbf{q}}^{\, \prime}} \, - \,
E_{ {\mathbf{q}} + {\mathbf{q}}^{\, \prime}}  }
{
E_{ {\mathbf{q}}^{\, \prime}} \, + \, E_{ {\mathbf{q}} + {\mathbf{ q}}^{\, \prime}}  \ + \
\sqrt{ ( E_{ {\mathbf{q}}^{\, \prime}} \, +
\, E_{ {\mathbf{ q}} + {\mathbf{ q}}^{\, \prime}} )^2
- {\mathbf{ q}}^{\ 2} }
}
\label{y1}
\end{eqnarray}
with $x=\hat q \cdot \hat q' $,
$y_2(q, q',x) = y_1(q', q,x)$ and $ E_{ {\mathbf{ q}} } \equiv \omega_m (
{\mathbf{ q}~} )$.

Proceeding as in ~\cite{bal,bkwp2,BBjpp} one gets 
 the following expression for the matrix element of the permutation operator P:
\begin{eqnarray}
_1\langle k ~, q ~ \alpha \vert ~ P ~ \vert k' ,~ q' ~ \alpha' \rangle_1 &=&
\int_{-1}^{1} dx { {\delta(k-\pi_1)} \over { k^{2} }  } ~
{ {\delta(k'-\pi_2)} \over { k'^{2} }  } ~  \cr
&~&
{1\over {N_1(q, q',x) } } ~ {1\over {N_2(q, q',x) } } ~
G_{\alpha \alpha'}^{BB} (q, q', x) ,
\label{eq20}
\end{eqnarray}
with
\begin{eqnarray}
\pi_1 &=& \sqrt{q'^2 +{1\over{4}}q^2(1+y_1)^2 + qq'x(1+y_1)}  \cr
\pi_2 &=& \sqrt{q^2 +{1\over{4}}q'^2(1+y_2)^2 + qq'x(1+y_2)} \cr
N_1(q, q',x) &\equiv&
N(\mathbf{q}~', -\mathbf{q} - \mathbf{q}~')  \cr
N_2(q, q',x) &\equiv&
N( -\mathbf{q} - \mathbf{q}~', \mathbf{q}~ ) ,
\label{eq21a}
\end{eqnarray}
and
\begin{eqnarray}
&&G^{BB}_{\alpha \alpha'}(q q' x) = {{4\pi}^{3/2}\over {2J+1}}
(-1)^{t'} \delta_{TT'}
\delta_{M_T M_{T'}} \sqrt{\hat t \hat t'} \left\{
\begin{array}{ccc}
1/2 & 1/2 & t \cr
1/2 &  T  & t' \cr
\end{array}
\right\} \sqrt{\hat {\lambda}} \cr
&~&
\sum_{\mu_2 \mu_3}  ( {1\over{2}} \mu_2 {1\over{2}} \mu_3  \vert s
 \mu_2 + \mu_3)  \cr
&~&
\sum_{\mu_2'} ( \sum_{m_3} 
{D^{1/2}_{m_3 \mu_3}}^*[R_w(B(-\mathbf{q}/2\omega_m(\mathbf{k})),
(-\mathbf{k},\omega_m(\mathbf{k}))]
{D^{1/2}_{m_3 \mu_2'}}[R_w(B(\mathbf{q}'/2\omega_m(\mathbf{k}')),
(\mathbf{k}',\omega_m(\mathbf{k}'))] ) \cr
&~&
\sum_{\mu_3'}  ( {1\over{2}}, \mu_2', {1\over{2}}, \mu_3'  \vert s',
 \mu_2' + \mu_3')  \cr
&~&
\sum_{m_1} ( \lambda 0 {1\over{2}}, m_1,  \vert I, m_1)
{D^{1/2}_{m_1 \mu_3'}}[R_w(B(-\mathbf{q}'/2\omega_m(\mathbf{k}')),
(-\mathbf{k}',\omega_m(\mathbf{k}'))]  \cr
&~&
\sum_{\mu}  ( l, \mu-\mu_2-\mu_3, s, \mu_2+\mu_3 \vert j ,\mu)
(-)^{\mu-\mu_2-\mu_3}Y_{l -(\mu-\mu_2-\mu_3}(\hat{\mathbf{p}})
( j, \mu, I, m_1 \vert J, \mu+m_1)
 \cr
&~&
\sum_{\mu'}  ( l', \mu'-\mu_2'-\mu_3', s', \mu_2'+\mu_3' \vert j', \mu')
Y_{l' \mu'-\mu_2'-\mu_3'}(\hat{\mathbf{p}}')
( j', \mu', I', \mu+m_1-\mu' \vert J, \mu+m_1)
 \cr
&~&
\sum_{m_2} ( \lambda', \mu+m_1-\mu'-m_2, {1\over{2}}, m_2  \vert I', \mu+m_1-\mu') \cr
&& ~~~~ {D^{1/2}_{m_2 \mu_2}}^*[R_w(B(-\mathbf{q}/2\omega_m(\mathbf{k})),
(\mathbf{k},\omega_m(\mathbf{k}))]
Y_{\lambda' \mu+m_1-\mu'-m_2}(\hat{\mathbf{q}}') .
\label{eq23}
\end{eqnarray}
We use standard notation $\hat l \equiv 2l+1$. It is assumed that
z-axis is along $\mathbf{q}$ and the momentum $\mathbf{q}~'$ lies in the x-z
plane what leads to the following components of the $\mathbf{q}$, 
$\mathbf{q}~'$, $\mathbf{k}$, and $\mathbf{k}~'$ vectors
\begin{eqnarray}
\mathbf{q} &=& [0, 0, q] \cr
\mathbf{q}~' &=& [q'\sqrt{1-x^2}, 0, q'x] \cr
\mathbf{k} &=& [q'\sqrt{1-x^2}, 0, q'x + {1\over {2}}q(1+y_1(q,q',x))] \cr
\mathbf{k}' &=& [-{1\over{2}}q'(1+y_2(q,q',x))\sqrt{1-x^2}, 0, 
-q-{1\over {2}}q'(1+y_2(q,q',x))x] 
\label{eq24}
\end{eqnarray}

Though the direct evaluation of Euler angles as arguments of 
the Wigner D-functions could be
used like in \cite{witrel}, here we use the $SL(2,\mathbb{C})$ 
 representations of Lorentz transfomations discussed in 
Appendix \ref{a1}. This leads to
\begin{eqnarray}
D^{1/2} [R_w(B(-\mathbf{q}/M_0),(\mathbf{k},m))] & = &  
B ( \mathbf{p}_2/m ) B ( - \mathbf{q}/M_0) B ( \mathbf{k}/m)\nonumber\\
&  = &  \sqrt{\frac{E_0 + M_0}{2 M_0}} \sqrt{\frac{\omega(k) + m}{\omega(p_2) + m}}\nonumber\\
&  - &  \frac{\mathbf{k} \cdot \mathbf{q} }{ \sqrt{2 M_0 ( E_0 + M_0 ) ( \omega(k) + m) ( \omega( p_2 ) +
m)}}\nonumber\\
&  + &  i \mathbf{k} \times \mathbf{q} \cdot \bsig \frac{1}{\sqrt{2 M_0 ( E_0 + M_0 )
 ( \omega(k) + m) ( \omega( p_2 ) + m)}}
\end{eqnarray}
where
\begin{eqnarray}
E_0 & = &  \sqrt{M_0^2 + q^2}\\
M_0 & = &  2 \omega(k)\\
\mathbf{p}_2 ( \mathbf{k},-\mathbf{q}) & = &  
\mathbf{k} - \mathbf{q} ( \frac{\omega(k)}{M_0} - 
\mathbf{k} \cdot \mathbf{q} \frac{1}{M_0 (
E_0 + M_0)} .
\end{eqnarray}

\end{document}